%% file: 2026snpd_tamada.tex
\documentclass[conference]{IEEEtran}
\IEEEoverridecommandlockouts

\usepackage{graphicx}
\usepackage{subcaption}
\usepackage{multirow}
\usepackage{url}
\def\BibTeX{{\rm B\kern-.05em{\sc i\kern-.025em b}\kern-.08em
    T\kern-.1667em\lower.7ex\hbox{E}\kern-.125emX}}

\newcommand{\bftt}[1]{\textbf{\texttt{#1}}}
\newcommand{\B}{B}

\begin{document}

\title{
    Cross-Platform Software Birthmarking for Real-World Binaries via Intermediate Representation\\
    \thanks{The part of this work was supported by JSPS Grant Number 26K14788.}
}

\author{\IEEEauthorblockN{Haruaki Tamada}
\IEEEauthorblockA{\textit{Kyoto Sangyo University.} \\
Kyoto, Japan\\
0000-0003-1838-9184}
}

\maketitle

\begin{abstract}
Software birthmarking detects plagiarism through characteristic program features,
yet cross-platform resilience remains under-evaluated.
This paper proposes a unified birthmarking approach for real-world binaries by lifting disparate formats into a common intermediate representation via Ghidra P-code.
Experiments across diverse platforms and languages demonstrate exceptional consistency across CPU architectures ($r=0.9846$),
independent of ISA (Instruction Set Architecture) specific details.
The study also identifies a ``dilution effect'' in Windows binaries,
in which the proliferation of library-derived functions degrades similarity scores.
Despite this noise, the Simpson index demonstrates superior discriminative power.
These findings clarify the practical capabilities and essential requirements for robust cross-platform birthmarking.
\end{abstract}


\section{Introduction}\label{sect:introduction}

Software theft is difficult to detect because stolen software is often modified to conceal plagiarism,
and identical source code can result in significantly different executables when compiled for different platforms.
To address this problem, software birthmarking has been proposed as a technique that extracts characteristic features from software and measures similarities between programs\cite{2005ieice_tamada,2019ajse_nazir}.
However, most existing birthmarking studies have focused on limited targets, such as Java bytecode,
and have not sufficiently evaluated executables across heterogeneous platforms, including different operating systems, binary formats, and CPU architectures.


The biggest problem with software birthmarking is that the target platforms in previous studies are limited to a few platforms.
Additionally, it has not been evaluated for the robustness of the birthmarking techniques across platforms and architectures.
In recent years, cross-compilation has become easier; we can compile a set of source codes and generate executables for various platforms.
It originates from the rise of compiler infrastructure such as LLVM\cite{2004cgo_lattner}.

Let's consider a scenario.
Suppose a developer creates proprietary software \textit{X} and releases it for macOS.
Unfortunately, \textit{X} is plagiarized, and an adversary releases it for a Windows platform after minor modifications.
Then, how can we detect this plagiarism?

Since the binary formats differ and the compilers used may also be different,
it is generally difficult to compare them.
The user should understand not only both binary formats,
but also the differences between them across platforms in analyzing plagiarism.
Furthermore, to date, no evaluation has been conducted of how different the binaries generated by cross-compilation actually are.

In this paper, we apply software birthmarking technology to executable files in real-world environments across multiple platforms and evaluate its performance.
Today, the most prominent executable formats include Windows PE\cite{windows_pe}, macOS Mach-O\cite{macos_macho}, and Linux ELF\cite{linux_elf}.
No unified method for handling these formats has been established, and even if birthmarking techniques exist for each format, significant effort is required to compare them.
On the other hand, LLVM uses a front-end mechanism to convert programs written in various languages into an intermediate language called LLVM IR, and then generates executable files for each platform using a back-end.
In other words, by constructing a birthmarking method targeting this intermediate language, we can achieve plagiarism detection that is independent of both platform and language.
Fortunately, a few binary lifting (decompilation) techniques have been proposed for converting executable files into a certain intermediate language.
By leveraging these, we can expect to develop platform- and language-independent birthmarking techniques that can target a wide range of executable files.

The rest of this paper is organized as follows.
Section~\ref{sect:related_work} reviews the related work.
Section~\ref{sect:preliminary} defines software birthmarks and describes their properties.
Section~\ref{sect:implementation} introduces the tool for the proposed birthmarking technique.
Section~\ref{sect:evaluation} presents the experimental setup and results.
Section~\ref{sect:discussion} discusses the results and the limitations of the proposed technique.
Finally, Section~\ref{sect:conclusion} concludes this paper.

\section{Related Work}\label{sect:related_work}

Software birthmarking was first proposed for Java programs to detect software theft without the need for embedding the watermarking~\cite{2005ieice_tamada}. 
Early research focused on defining various types of birthmarks, such as those based on instruction sequences ($k$-grams)~\cite{2005sac_myles}. 
While these foundational works established the properties of credibility and resilience, they were primarily designed for high-level languages or specific instruction set architectures (ISAs). 
In contrast, our work aims to provide a cross-architecture solution by leveraging intermediate representations.

Binary code similarity has become a major research area, with applications in vulnerability discovery and malware analysis~\cite{2021surv_haq}. 
Recent studies have employed binary lifting to translate diverse ISAs into a common intermediate representation (IR) to mitigate architecture-specific differences. 
Tools like discovRE~\cite{2016ndss_eschweiler} and BinGo~\cite{2016fse_chandramohan} use structural and semantic features from IR to identify similar code across platforms. 
Furthermore, state-of-the-art methods such as HermesSim~\ utilize Ghidra P-code to build semantics-oriented graphs for high-precision matching via graph neural networks\cite{2024usenix_he_haojie}.
Our approach differs by using a training-free, deterministic matching algorithm that prioritizes explainability in the context of birthmarking.

To handle the ever-increasing volume of software, scalable search-based detection systems have been developed. 
Mituba applies search engine technologies like inverted indexing to scale up software theft detection across millions of projects\cite{2018icsim_nakamura}. 
Similarly, Ghidra's BSim uses locality-sensitive hashing on P-code features for rapid similarity lookups\cite{ghidra_bsim}. 
While these systems are excellent for large-scale screening, our proposed birthmarking technique provides a more rigorous,
second-stage identification process that cross-checks suspected pairs across different architectures with high fidelity.

\section{Software Birthmarks}\label{sect:preliminary}

\subsection{Definitions of software birthmarks}\label{sect:definition}

A software birthmark is a collection of unique characteristics of a given software, which can be used to identify the software.
It is used to detect software theft, plagiarism, and tampering.

Let $\B(p)$ be a function that extracts birthmarks from software $p$.
Here, the birthmark can be extracted either statically or dynamically from $p$ (possibly with a certain input $I$ to $p$ for dynamic birthmarks).
Then, two birthmarks $\B(p)$ and $\B(q)$ provide a method for calculating the similarity, which reflects the similarity between the original software $p$ and $q$.
The similarity is a value between $0$ and $1$, where $1$ means that the two software are identical, and $0$ means that they are completely different.

Hopefully, a software birthmark should satisfy the following two properties:
\begin{itemize}
\item \textbf{Credibility} means that the similarity between two software programs developed independently should be low, even if they have identical functions, and
\item \textbf{Resilience} indicates that the similarity between an altered program and the original should remain high.
\end{itemize}

\subsection{The workflow of birthmark-based theft detection}\label{sect:workflows}

The primary purpose of software birthmarking is to detect software that is quite similar to the original software from the enormous number of software sets.
To achieve this, the following steps should typically be employed.

\begin{itemize}
\item \textbf{Preparation} Gather the set of software to be examined,
\item \textbf{Extraction}  Extract birthmarks from each software,
\item \textbf{Comparison}  Compare the extracted birthmarks, and
\item \textbf{Detection}   Analyze the comparison results and identify software that may be stolen.
\end{itemize}

\subsubsection{Preparation step}\label{sect:preparation}

In the preparation step, we gather the set of software to be examined and compare them with the original software in the following steps.
The software in the set is sourced from various sources, including software repositories and websites.
If we have a specific suspicion of plagiarized software, the set will be smaller.
However, if we are conducting a broad-based plagiarism check, a vast number of sets will be required.

\subsubsection{Extraction step}\label{sect:extraction}

In this step, we extract birthmarks from each software in the set and the original software.
There are various types of birthmarks, such as opcode sequences, $k$-gram based opcodes, etc.
However, the choice of which type of birthmarks to select is left to the user.
Furthermore, since the performance of birthmark types has not yet been sufficiently investigated, a comprehensive study is needed in the future.

\subsubsection{Comparison step}\label{sect:comparison}


Generally, software consists of multiple modules, each containing several functions.
Then, the unit for birthmark extraction is typically a function unit.
Therefore, the comparison of birthmarks should be performed at the function unit level, and the similarities within each function unit should be aggregated to calculate the overall similarity of the software.

Although comparison algorithms for function units have been discussed in previous studies,
methods for calculating the similarity of the entire software have received less attention.
In the earliest studies, a simple average was used\cite{2005ieice_tamada},
however, the average was pointed out to underestimate the similarity, and top-$n$ selection was proposed by Fedorov et al.\cite{2024wsse_fedorov}.
Other methods include using combinatorial optimization algorithms such as the Hungarian algorithm\cite{1955hungarian_kuhn,1987lapjv_jonker,2018ijndc_yokoi}.

Here, we introduce the comparison algorithms used in previous research.
For the definition of the birthmark comparison algorithms,
let $\B(p)$ and $\B(q)$ be the birthmarks which contains sequences of elements extracted from two software $p$ and $q$, respectively.
Then, the comparison algorithms calculate the similarity between $p$ and $q$ based on the similarities between the birthmarks in $\B(p) = \{ e_{p1}, e_{p2}, \ldots, e_{pm_p} \}$ and $\B(q) = \{ e_{q1}, e_{q2}, \ldots, e_{qm_q} \}$.
Also, let $\mathcal{F}(\B(p))$ be the function by vectorize the birthmark $\B(p)$ by its frequencies, and
$f_p(i)$ and $f_q(i)$ be the frequency values of $e_i$ in $\B(p)$ and $\B(q)$, respectively.
$\mathcal{S}(\B(p))$ be the function that extracts the set of elements in the birthmark $\B(p)$.
Then, the comparison algorithms are defined as shown in Table~\ref{table:comparison_algorithms}.

\begin{table}[tb]
    \centering
    \begingroup
    \renewcommand{\arraystretch}{2.5}
    \caption{Comparison algorithms for birthmark similarity calculation}
    \label{table:comparison_algorithms}
    {\footnotesize
    \begin{tabular}{p{0.35\linewidth} l} 
\textbf{Dice index}             & $\displaystyle \frac{2|\mathcal{S}(\B(p)) \cap \mathcal{S}(\B(q))|}{|\mathcal{S}(\B(p))| + |\mathcal{S}(\B(q))|}$ \\
\textbf{Jaccard index}          & $\displaystyle \frac{|\mathcal{S}(\B(p)) \cap \mathcal{S}(\B(q))|}{|\mathcal{S}(\B(p)) \cup \mathcal{S}(\B(q))|}$ \\
\textbf{Simpson index}          & $\displaystyle \frac{|\mathcal{S}(\B(p)) \cap \mathcal{S}(\B(q))|}{\min(|\mathcal{S}(\B(p))|, |\mathcal{S}(\B(q))|)}$ \\
\textbf{LCS (Longest Common Subsequence)} & $\displaystyle \frac{|\mathrm{lcs}(\B(p), \B(q))|}{\max(|\B(p)|, |\B(q)|)}$ \\
\textbf{Levenshtein similarity} & $\displaystyle 1 - \frac{\mathrm{levenshtein}(\B(p), \B(q))}{\max(|\B(p)|, |\B(q)|)}$ \\
\textbf{Euclidean similarity}   & $\displaystyle 1 - \frac{\sqrt{\sum(f_p(i) - f_q(i))^2}}{\sqrt{\sum f_p(i)^2} + \sqrt{\sum f_q(i)^2}}$ \\
\textbf{Cosine similarity}      & $\displaystyle \frac{\mathcal{F}(\B(p)) \cdot \mathcal{F}(\B(q))}{\|\mathcal{F}(\B(p))\| \|\mathcal{F}(\B(q))\|}$ \\
\textbf{Weighted Jaccard index} & $\displaystyle \frac{\sum_{i=1}^{n} \min(f_p(i), f_q(i))}{\sum_{i=1}^{n} \max(f_p(i), f_q(i))}$ \\
    \end{tabular}}
    \endgroup
\end{table}


\subsubsection{Detection step}\label{sect:detection}

Finally, we analyze the comparison results and identify software that may be stolen.
The previous studies introduced the threshold-based method, which identifies software as stolen if the similarity exceeds a certain threshold (typical threshold value is $\varepsilon = 0.75$).
%
Note that this paper does not consider such detection methods,
since the main focus of this work is the similarity distributions of executables and
the relationships between similarities and software characteristics, such as programming language, compiler, and features.

\section{Implementation}\label{sect:implementation}

\subsection{\textbf{Oinkie}: The birthmarking toolkit}\label{sect:oinkie}

The author has developed \textbf{Oinkie}, a software birthmarking toolkit,
which can extract several types of birthmarks from Oinkie-IR (Intermediate Representation) and compare them\footnotemark.
Oinkie-IR is a simple JSON format, which includes a list of instructions (opcodes and their operands) for every function in an executable file.

In the current version of \textbf{oinkie}, we can extract three types of birthmarks:
opcode sequence (\texttt{opseq}), opcode frequency (\texttt{opfreq}), and opcode set (\texttt{opset}).
Also it provides the comparison algorithm shown in Table~\ref{table:comparison_algorithms}.

\subsection{Similarity calculation between two executables}\label{sect:similarity}

The algorithms shown in Table~\ref{table:comparison_algorithms} calculate the similarities between two functions in the executables.
However, the birthmarking scenario requires the similarity between two executables.
Therefore, we should aggregate the similarity matrix among functions by some method.
This paper employs the aggregation method top-$n$ selection \cite{2024wsse_fedorov}, and
hungarian algorithm~\cite{1955hungarian_kuhn,1987lapjv_jonker}.

Top-$n$ selection finds the max similarities in each row and column of the similarity matrix, and
calculates the average of the selected similarities.
Hungarian algorithm is a combinatorial optimization algorithm,
which finds the optimal matching between two sets of functions in the executables.
In this paper, we use top-$n$ selection with $n=1$ and $n=\infty$ (all) for the evaluation.

\subsection{Binary lifting}\label{sect:binary_lifting}

To apply the proposed birthmarking techniques to executable files, we firstly should convert the executables to oinkie-IR.
Today, there are several intermediate representations for executable files, such as
LLVM-IR/BC\footnotemark,
Ghidra P-code\footnotemark,
IDA Pro microcode\footnotemark,
Binary Ninja IL (Intermediate Language)\footnotemark, and etc.
In this paper, we use Ghidra P-code as the intermediate representation for executable files,
since it is open-source and supports various architectures and platforms.
Also, Ghidra supports the binary lifting (decompilation) of executable files into P-code, and
thus we can easily convert executable files into the intermediate representation.
Then, oinkie-IR is generated from the obtained P-code using a simple converter that extracts the symbol names,
opcodes, and their operands, and stores them into a JSON file.

\footnotetext[1]{\url{https://github.com/tamada/oinkie}}
\footnotetext[2]{\url{https://llvm.org/docs/LangRef.html}}
\footnotetext[3]{\url{https://ghidra.re/ghidra_docs/languages/html/pcoderef.html}}
\footnotetext[4]{\url{https://hex-rays.com/blog/microcode-in-pictures}}
\footnotetext[5]{\url{https://docs.binary.ninja/dev/bnil-overview.html}}

\section{Evaluation}\label{sect:evaluation}

\subsection{Setup}\label{sect:setup}

Through the following experiments, we evaluate the birthmarks by the following steps.

\begin{enumerate}
\item obtain the target software binaries (such as Mach-O, PE, and/or ELF),
\item lift the obtained binaries into P-code using Ghidra,
\item convert the obtained P-code into oinkie-IR,
\item extract opseq, opfreq, or opset birthmarks from the obtained oinkie-IR, and
\item compare the extracted birthmarks and calculate the similarities.
\end{enumerate}

\subsection{Resemblance Evaluation}\label{sect:experiment0}

\subsubsection{Target software}\label{sect:target_exp1}

\begin{figure}[tb]
    \centering
    \includegraphics[width=0.95\linewidth]{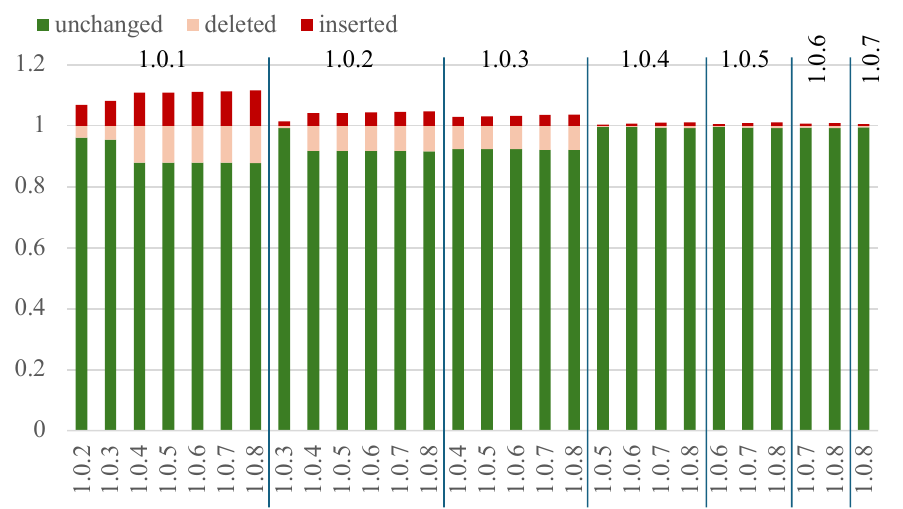}
    \caption{The differences between the versions of \bftt{bzip2}}\label{fig:diff_exp1}
\end{figure}

In the first experiment, we compare the birthmarks of the essentially near-identical software to confirm that they exhibit high similarity.
This evaluation is crucial for establishing a baseline before assessing the similarities of different software in subsequent experiments. 
By ``essentially identical,'' we refer to software that shares the same OS, architecture, programming language, and compiler, with only slight modifications.
For this purpose, we utilize binaries from different versions of \bftt{bzip2} (versions 1.0.1 to 1.0.8), a widely used data compression utility.
These binaries were obtained from the official website\footnote{\url{https://sourceware.org/pub/bzip2/}}.
The differences of the source code between these versions are shown in Figure~\ref{fig:diff_exp1} with stacked chart, calculated by \texttt{difflib} in Python\footnote{\url{https://docs.python.org/ja/3/library/difflib.html}}.
The horizontal axis represents the compared version pairs, grouped by the base version.
The bars show the ratio of unchanged, deleted, and inserted lines relative to the total lines of the older version.
From figure~\ref{fig:diff_exp1}, we can see that the differences among versions are quite small,
with only a few percent of code added, and removed.
%


Next, we compile these versions of \bftt{bzip2} from source code into arm64 Mach-O executables on macOS using \texttt{clang} without any compile options.
Then, we applied the proposed birthmarking techniques to calculate the similarities among these versions.
Since the updates between these versions mainly involve bug fixes and minor improvements,
their core logic remains largely unchanged, making them ideal for evaluating the resemblance of birthmarks.

\subsubsection{Results}\label{sect:results_exp0}

\begin{figure}[tb]
  \centering
  \includegraphics[width=0.95\linewidth]{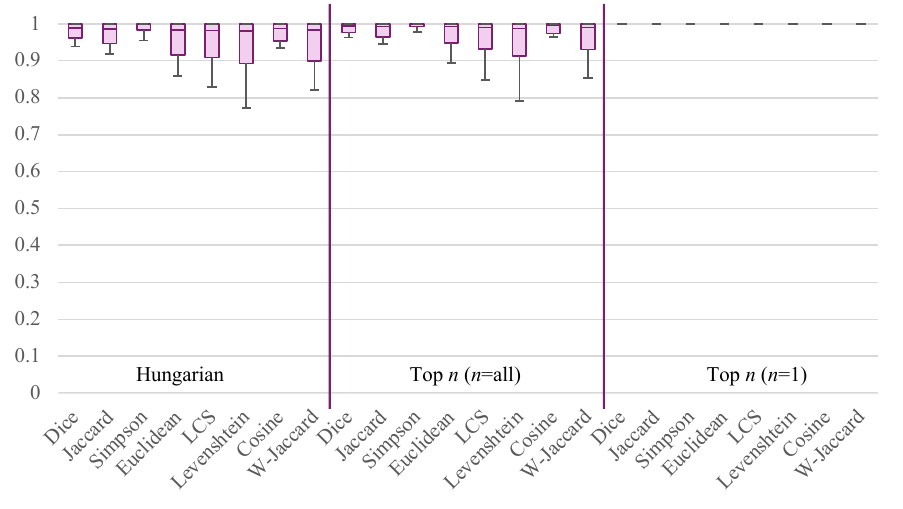}
  \caption{
    Similarities among the different versions of \bftt{bzip2}
}
  \label{fig:boxplot_exp1}
\end{figure}

Figure~\ref{fig:boxplot_exp1} shows the boxplot of similarities among the different versions of \bftt{bzip2}.
The horizontal axis represents the different comparison algorithms, and the vertical axis represents the similarity values.
The results show that the similarities among the different versions of \bftt{bzip2} are quite high, with all values exceeding 0.75 and many close to 1.0.
In the results, the lowest similarity is 0.77 between version 1.0.1 and 1.0.3 by Levenshtein similarity (Hungarian and top-$n$ ($n=\infty$)).
Although this similarity is lower than the other comparison results, it is still sufficiently high to be considered as the same software.

Comparing the three aggregation methods, the Hungarian algorithm provided the most discriminative results,
capturing subtle differences between versions that were sometimes overlooked by the top-$n$ methods.
Given its mathematical rigor and sensitivity, we will focus on the results obtained via the Hungarian algorithm in the subsequent experiments.

\subsection{Credibility evaluation}\label{sect:experiment1}

\subsubsection{Target software}\label{sect:target_exp2}

This experiment evaluates differences of the birthmarks from the different software.
For this purpose, we should choose software with the same purpose and different authors.
Therefore, we first selected \textbf{bzip2} as the target software\footnotemark.
We also chose a Go implementation of bzip2 (\textbf{bzip2go})\footnotemark.
In addition, the author implemented a Rust version of bzip2 (\textbf{bzip2rs}) because none existed\footnotemark.

On the other hand, we evaluate differences in birthmarks across programming language levels.
For this, we should use the same algorithm in different languages.
Therefore, we requested generative AI, Google Gemini to implement factorization, MD5, and SHA256 in C, Go, and Rust.
The prompt to generative AI was "Write the \textit{SPEC} logic in \textit{LANGUAGE} without standard libraries."
The italic words were replaced with the corresponding specification and language.
Of course, there are many other famous programming languages.
However, we chose C, Go, and Rust, since they can generate executables and are easy to implement.

Moreover, the algorithms of \bftt{bzip2}, \bftt{md5}, and \bftt{sha256} use many bit calculations,
and thus are expected to have similar instruction sets and high similarities.
On the other hand, the algorithm of \bftt{factorization} is implemented by trial division,
which uses multiplication and remainder calculations, and no bit calculations.
Therefore, it is expected to have a different instruction set and low similarities from the other three software.
Next, we compile the obtained four software in three languages to arm64 Mach-O executables on macOS.
We employ \texttt{clang} for C, \texttt{go} for Go, and \texttt{rustc} for Rust.
No compile options were specified in any cases.
Finally, we obtain 12 executables shown in Table~\ref{table:target_exp2}.
Then, we apply the proposed birthmarking techniques to calculate the similarities among them.

\footnotetext[8]{\url{https://sourceware.org/git/?p=bzip2.git;a=summary} (hash: af79253)}
\footnotetext[9]{\url{https://github.com/pedroalbanese/bzip2} (hash: 575eca0)}
\footnotetext[10]{\url{https://github.com/tamada/bzip2rs} (hash: 6972810)}

\begin{table}[tb]
    \centering
    \caption{The target software for the credibility evaluation}\label{table:target_exp2}
    {\footnotesize
    \begin{tabular}{l l | l l}\hline \hline
\textbf{Label} & \textbf{Language} & \textbf{Label} & \textbf{Language} \\ \hline
\bftt{bzip2\_gcc}     & C    & \bftt{sha256\_gcc}    & C    \\
\bftt{bzip2\_go}      & Go   & \bftt{sha256\_go}     & Go   \\
\bftt{bzip2\_rs}      & Rust & \bftt{sha256\_rs}     & Rust \\ \hline
\bftt{md5\_gcc}       & C    & \bftt{factorize\_gcc} & C    \\
\bftt{md5\_go}        & Go   & \bftt{factorize\_go}  & Go   \\
\bftt{md5\_rs}        & Rust & \bftt{factorize\_rs}  & Rust \\ \hline
    \end{tabular}}
\end{table}

\subsubsection{Result}\label{sect:result-credibility}

\input{inputs/experiment2/heatmap}

\begin{figure}[tb]
  \centering
  \includegraphics[width=0.95\linewidth]{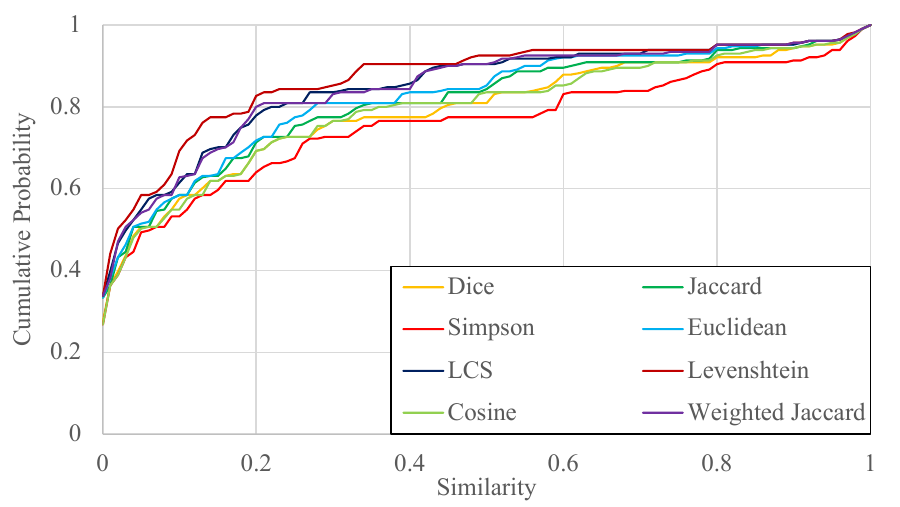}
  \caption{
    ECDF of similarities between independent software.
  }
  \label{fig:credibility-ecdf}
\end{figure}

Figure~\ref{fig:credibility-hungarian} shows the similarity distributions as the heatmap among the targets shown in Table~\ref{table:target_exp2}.
The horizontal axis and vertical axis of each figure represent the target shown in Table~\ref{table:target_exp2}, and
the color of each cell is represented in Figure \ref{fig:heatmap-scaler}, which shows the similarity between the two targets.

From Figure~\ref{fig:credibility-hungarian}, the diagonal cells are all red, representing identical software (similarity $1$).
In contrast, red parts in the middle of Figure~\ref{fig:credibility-hungarian} correspond to the pairs of \bftt{md5\_gcc} and \bftt{sha256\_gcc}.
This is because both algorithms involve many bit calculations and have similar opcode sets.

Also, Figure~\ref{fig:credibility-ecdf} shows the chart of ECDF (Empirical Cumulative Distribution Function).
The horizontal and vertical axes represent the similarity and the cumulative probability, respectively.
From the figure, all of lines rise to about 0.8 under 0.4 similarity, which means that about 80\% of the pairs have similarity less than 0.4.
This result shows that the proposed birthmarks can capture the differences between different software, and thus the similarities between different software are quite low.



\subsection{Resilience evaluation}\label{sect:experiment3}

\subsubsection{Target software}\label{sect:target_exp3}

\input{inputs/experiment3/targets_with_compilers}

This experiment evaluates the similarities in the birthmarks produced by the same software compiled for different architectures and platforms.
We employ \bftt{bzip2} and its Go and Rust implementations as the target software, which is the same software as in Section~\ref{sect:target_exp2}.
We compile \bftt{bzip2} for macOS Mach-O, Linux ELF, and Windows PE in amd64 and arm64 architectures.
Note that the number of functions in the source code is: 129 (\bftt{bzip2}), 5 (\bftt{bzip2go}), and 22 (\bftt{bzip2rs}) via \texttt{ctags}.

We prepare the executables using \texttt{clang} and \texttt{gcc} for C, \texttt{go} and \texttt{TinyGo}\footnotemark for Go, and \texttt{rustc} for Rust.
Note that \bftt{bzip2rs} has two features for compression: a pure Rust implementation and a delegate to \texttt{libbz2}.
We chose one at compile time.
In the Windows environment, we also use \texttt{msvc} (Microsoft Visual C++) for \bftt{bzip2}.
Almost all compilations were performed in GitHub Actions\footnotemark.
Also, we use Docker images (golang:1.26.3-bookworm and tinygo/tinygo:0.40.0) to compile \bftt{bzip2go} across all three platforms.
Compiling \bftt{bzip2rs} was performed on the local macOS machine with zigbuild for linking\footnotemark.
Specifically, for the Windows platform, we use xwin\footnotemark for \bftt{bzip2rs} on macOS,
with target pc-windows-msvc for amd64 and arm64 architectures.
Unfortunately, we could not compile \bftt{bzip2go} with \texttt{TinyGo} on Windows platform because it does not support Windows well.
In addition, in macOS environment, \texttt{gcc} is actually \texttt{clang}.
Therefore, we should install it explicitly to use the actual \texttt{gcc}\footnotemark.

The resulting 36 executables are shown in Table~\ref{table:target_exp3}, which
also includes their file sizes, the compilers used, their versions, and the number of functions in the Oinkie-IR (birthmarks).
The number of \bftt{bzip2} functions is reduced after compiling and lifting, which may cause inline expansion or unused functions.

\footnotetext[11]{\url{https://tinygo.org}}
\footnotetext[12]{\url{https://github.com/tamada/2026snpd_tamada_experiments}}
\footnotetext[13]{\url{https://github.com/rust-cross/cargo-zigbuild}}
\footnotetext[14]{\url{https://github.com/rust-cross/cargo-xwin}}
\footnotetext[15]{see the version information of \texttt{gcc} on macOS environment.}


Besides, \texttt{clang}, \texttt{TinyGo}, and \texttt{rustc} use LLVM,
while \texttt{gcc}, \texttt{msvc}, and \texttt{go} rely on distinct backend technologies.
This is reflected in the file sizes in Table~\ref{table:target_exp3}.
Go's feature-rich runtime yields the largest executables, followed by TinyGo's garbage-collected runtime.
Rust's thin runtime and C's low-level nature result in smaller binaries.
%

Ideally, the compiler versions should be the same across all platforms, however, due to the complexity of preparing, we could not achieve that.
Then, we apply the proposed birthmarking techniques to calculate their similarities.




\subsubsection{Result I: Cross-architecture}\label{sect:cross-architecture}

\begin{figure}[bt]
    \centering
    \includegraphics[width=0.95\linewidth]{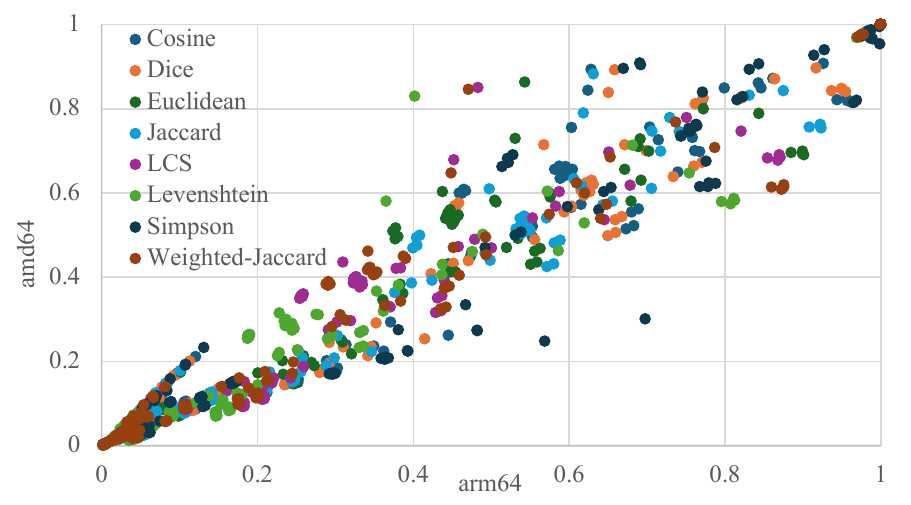}
    \caption{Scatter plot of similarities between different architectures ($r=0.9846$).}
    \label{fig:scatter}
\end{figure}


First, we evaluated the impact of architectural differences by examining whether relative similarity rankings are preserved across architectures (i.e., whether a pair with high similarity on amd64 also shows high similarity on arm64).
Figure~\ref{fig:scatter} shows the correlation between these similarities;
the horizontal axis represents amd64 pairs, and the vertical axis represents the corresponding arm64 pairs (correlation coefficient $r = 0.9846, p < 0.0001$).
This result suggests that the proposed birthmark focuses on aspects closer to the essence of the algorithm,
such as data flow and types of operations, rather than physical operations like register names or stack manipulations.
However, this may also depend on the decompilation performance of Ghidra.

\subsubsection{Result II: Analysis of Platform-Specific Anomalies}\label{sect:windows-pe-clang}

\begin{figure}[bt]
    \centering
    \includegraphics[width=0.95\linewidth]{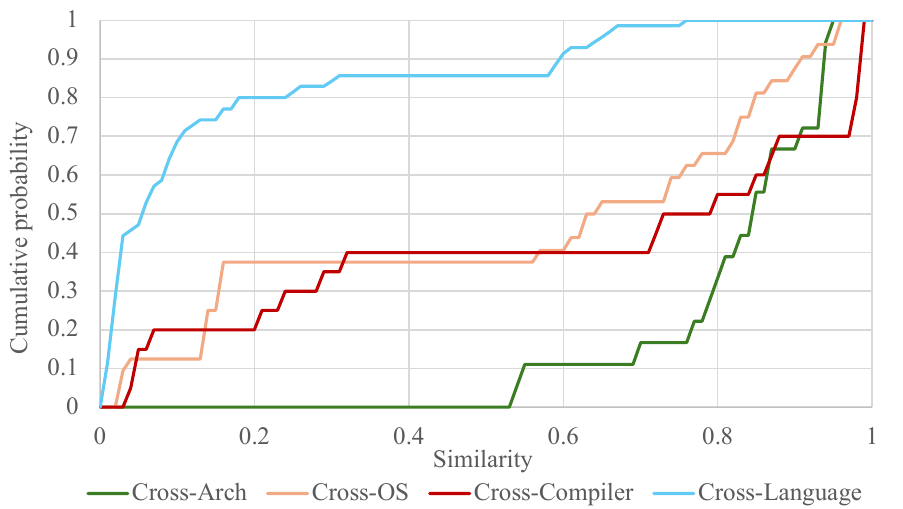}
    \caption{ECDF of cross-factor using cosine similarities.}\label{fig:ecdf2}
\end{figure}

To examine the overall trends, Figure \ref{fig:ecdf2} shows ECDF plots for cross-arch, cross-os, cross-compiler, and cross-language,
using cosine similarity to normalize for instruction count variations,
focusing on the absolute similarity levels rather than the relative rankings examined in Section \ref{sect:cross-architecture}.
The horizontal axis in Figure \ref{fig:ecdf2} represents similarity, and the vertical axis represents the cumulative proportion up to that similarity.
Furthermore, cross-X uses the average of pairs where only X differs from the overall comparison.

From Figure \ref{fig:ecdf2}, Cross-Arch rises after 0.5 and increases sharply after 0.75.
This indicates that the minimum similarity for Cross-Arch is 0.5, and that most cases have a similarity of 0.75 or higher.

On the other hand, Cross-OS and Cross-Compiler exhibited an anomalous bimodal distribution, with similarity rising sharply at both 0.2 and 0.7.
Upon investigating this ``gap,'' it was found that the group with low similarity consisted entirely of comparisons involving the Windows platform.
For example, the similarity for \texttt{clang} (Linux vs. Win) was a mere 0.03.
The root cause lies in the explosive increase in the number of functions, shown in Table \ref{table:target_exp3}.
The Windows version of \texttt{clang} contained 2,098 functions, whereas the Linux version had only 68.
This surge in functions, caused by the static linking of the C runtime library and system stubs,
triggered a severe ``dilution effect'' in the Hungarian method, causing bzip2's core logic to be buried under thousands of non-matching library functions.


This bimodal pattern also extends to Cross-Compiler comparisons restricted to the Windows platform (e.g., Windows-\texttt{clang} vs. Windows-\texttt{gcc}),
since \texttt{clang}'s extensive function inflation dilutes its birthmark regardless of the comparison target's origin.
Furthermore, this noise caused an unexpected reversal: Windows-\texttt{clang} exhibited a higher similarity (0.74) to Go (Linux) than its own Linux-\texttt{clang} counterpart.
This reversal can be explained by the comparable degree of dilution—both Windows-\texttt{clang} (2,098 functions) and Linux-Go (2,572 functions) contain a similarly large number of functions,
whereas Linux-\texttt{clang} (68 functions) does not, so the apparent similarity likely reflects comparable noise levels rather than genuine logical commonality.
This suggests that environmental noise can override the characteristics of compilers and languages, implying that in certain environments, noise-robust aggregation methods such as top-$n$ selection are necessary.

\begin{figure}[bt]
    \centering
    \includegraphics[width=0.95\linewidth]{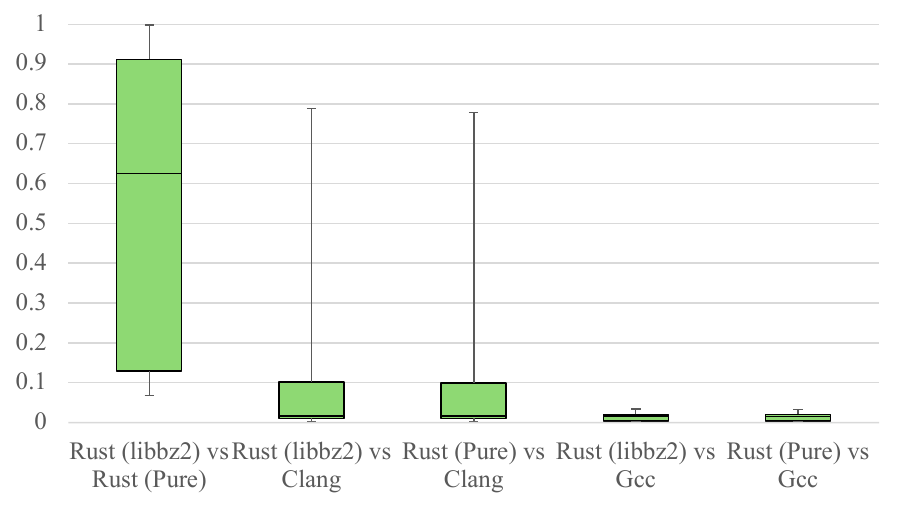}
    \caption{Box plot of the similarities among Rust and C implementations}
    \label{fig:rust_box_plot}
\end{figure}

\input{inputs/experiment3/scatter}

\subsubsection{Result III: Analysis of Language Barrier (Rust vs. C)}\label{sect:cross-languages}

As shown in the ECDF (Fig. \ref{fig:ecdf2}),
the ``Cross-Language'' category exhibits the lowest similarity among all categories,
indicating that the choice of programming language is the most dominant factor in birthmark variation.
To evaluate this barrier, we analyzed the similarities between C and Rust implementations using four compiler/feature sets:
\texttt{rustc} (lib), \texttt{rustc} (pure), \texttt{clang}, and \texttt{gcc}.

The results are shown in Fig. \ref{fig:rust_box_plot}.
As predicted by the ECDF, inter-language similarities remained consistently low, ranging from 0.1 to 0.3.
Notably, even the Rust implementation utilizing the C-based \texttt{libbz2} library failed to show high similarity to the original C binaries.
This confirms that the distinct runtime environments and memory management mechanisms of each language dominate the resulting binary structure,
overriding the underlying logical commonality.
While intra-language similarities remain high (median was approximately 0.626), overcoming the cross-language barrier remains a significant challenge.

\subsubsection{Result IV: Comparative Analysis of Comparison Algorithms}\label{sect:algorithm}

To evaluate the robustness and discriminative power of the eight comparison algorithms,
we visualized the similarities of inter-OS pairs (horizontal axis) and inter-compiler pairs (vertical axis) as scatter plots in Fig. \ref{fig:resilience-scatter}.
In these plots, the upper-right quadrant represents high resilience to both operating system and compiler variations.

Rust (diamonds) is distributed along the top of the plots.
It forms two distinct clusters: one in the upper-right (Unix-to-Unix comparisons) and another in the upper-left (Windows-to-Unix comparisons),
indicating high compiler resilience but persistent OS sensitivity.
Go (triangles) is clustered in the lower-right quadrant, showing high stability across OSes but weakness against compiler differences (Go vs. TinyGo).
In contrast to Rust and Go, C (circles) is distributed across all four quadrants.
This diversity indicates that C birthmarks are highly sensitive to the specific combination of OS and compiler;
while Unix-based C binaries show high resilience in the upper-right,
any comparison involving the Windows environment inevitably falls into the lower-left corner due to the dilution effect.


Set-based methods (Dice, Jaccard, Simpson) and Cosine similarity exhibited more pronounced patterns than others,
preserving clear clusters by focusing on logical orientation or set overlap;
Cosine similarity was notably resilient to proportional changes in instruction counts.


In contrast, sequence-based methods (Euclidean, LCS, Levenshtein) and Weighted Jaccard yielded more conservative,
lower scores, being highly sensitive to instruction order or frequency variations across compiler backends.
Among all tested algorithms, the Simpson index demonstrated the highest discriminative power,
with dots most clearly clustered toward the corners of the plot.

\section{Discussion}\label{sect:discussion}

\subsection{Computational Efficiency and Algorithm Selection}\label{sect:efficiency}

The experimental results highlight a significant trade-off between the precision of comparison algorithms and their computational overhead. 
We conducted the experiments on an Apple MacBook Air M5 with 32GB of unified memory (macOS 26.5.1).
As described in Section~\ref{sect:setup}, the number of comparisons increases quadratically with the number of binaries ($_{n}C_{2} + n$).
For the robustness evaluation (36 binaries, 666 comparisons), set-based algorithms completed within approximately 8 hours. 
In contrast, sequence-based algorithms like Levenshtein and LCS required approximately 56 hours.
This significantly higher cost is due to their $O(L^2)$ complexity per pair, where $L$ denotes the maximum opcode sequence length.

Despite the higher cost, sequence-based methods did not show a proportional increase in discriminative power for our Oinkie-IR-based birthmarks. 
In fact, set-based methods such as Simpson and Jaccard demonstrated a clearer separation between similar and dissimilar pairs. 
For practical cases, we recommend a tiered approach: employing set-based methods for rapid initial screening and reserving sequence-based methods for deep verification of high-similarity candidates.

\subsection{Cross-Platform Robustness and Practical Implications}\label{sect:robustness_disc}

The high correlation coefficient ($0.9846$) observed in our cross-architecture experiments (Section~\ref{sect:cross-architecture}) provides empirical evidence that Oinkie-IR-based birthmarking effectively abstracts away low-level ISA differences. 
By focusing on data flow and operation types in Ghidra's P-code, the birthmarks capture the software's underlying logic. 
This resilience suggests that our method is viable for real-world scenarios, such as detecting software plagiarism or license violations across different operating systems and CPU architectures.

\subsection{Threats to Validity}\label{sect:threats}

\subsubsection{Internal Validity} 

A potential threat lies in our use of AI-generated code (from Google Gemini) as a proxy for independent software in the credibility evaluation. 
There is a concern that LLMs might produce similar code for identical prompts, potentially inflating similarity scores. 
However, our results consistently showed low similarity between AI-generated samples and original implementations, indicating that the birthmarks are sensitive to implementation-level differences even when functional requirements are identical.

\subsubsection{External Validity} 

Our evaluation was limited to the \bftt{bzip2} utility and its variants. 
While \bftt{bzip2} is a standard benchmark for birthmarking, the results may not generalize to GUI-heavy applications, kernel-level code, or software with heavy obfuscation. 

\subsubsection{Construct Validity} 

The stability of the birthmarks depends on the quality of the binary lifting and the environment-specific binary structure.
The observed drop in similarity when using \texttt{clang} on Windows (Section \ref{sect:windows-pe-clang}) is primarily attributed to the proliferation of library-derived functions,
which dilute the logical birthmarks of the target software.
This indicates that the current birthmarking process is sensitive to the surrounding environment and runtime libraries.
Further investigation into normalizing P-code and implementing automated library code filtering is required to enhance the resilience of the Oinkie-IR against such environmental noise.


\section{Conclusion}\label{sect:conclusion}

This paper evaluated IR-based cross-platform birthmarking via binary lifting.
Experiments confirm that the method provides high similarity for identical logic and exceptional consistency across CPU architectures ($r=0.9846$),
abstracting away ISA-specific details. While programming language remains a significant barrier,
we identified that environmental noise on platforms like Windows causes a ``dilution effect'' that degrades similarities.
Analysis of eight algorithms shows that set-based methods, especially the Simpson index, offer superior discriminative power by isolating logical commonality from such noise.
While highly effective for cross-architecture analysis, IR-based birthmarking requires further improvements in cross-language robustness.
Future work will focus on automated library filtering to mitigate environmental impacts and optimizing similarity computations for enhanced scalability.

\footnotesize
\bibliographystyle{IEEEtran}
\bibliography{2026snpd_tamada}

\end{document}

%% file: inputs/experiment2/heatmap.tex
\begin{figure}[tb]
  \centering
    \begin{subcaptionblock}{0.24\linewidth}
      \includegraphics[width=\linewidth]{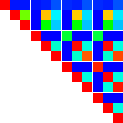}
      \caption{Dice}
    \end{subcaptionblock}
    \begin{subcaptionblock}{0.24\linewidth}
      \includegraphics[width=\columnwidth]{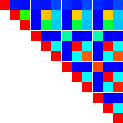}
      \caption{Jaccard}
    \end{subcaptionblock}
    \begin{subcaptionblock}{0.24\linewidth}
      \includegraphics[width=\columnwidth]{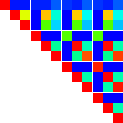}
      \caption{Simpson}
    \end{subcaptionblock}
    \begin{subcaptionblock}{0.24\linewidth}
      \includegraphics[width=\columnwidth]{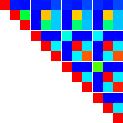}
      \caption{Euclidean}
    \end{subcaptionblock}

    \vspace{0.05cm}

    \begin{subcaptionblock}{0.24\linewidth}
      \includegraphics[width=\linewidth]{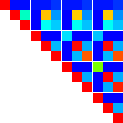}
      \caption{LCS}
    \end{subcaptionblock}
    \begin{subcaptionblock}{0.24\linewidth}
      \includegraphics[width=\linewidth]{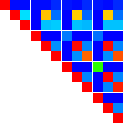}
      \caption{Levenshtein}
    \end{subcaptionblock}
    \begin{subcaptionblock}{0.24\linewidth}
      \includegraphics[width=\linewidth]{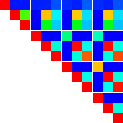}
      \caption{Cosine}
    \end{subcaptionblock}
    \begin{subcaptionblock}{0.24\linewidth}
      \includegraphics[width=\linewidth]{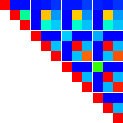}
      \caption{W-Jaccard}
    \end{subcaptionblock}
  \caption{The similarities among the targets shown in Table~\ref{table:target_exp2} (Hungarian method)}\label{fig:credibility-hungarian}

  \vspace{-0.05cm}

  \includegraphics[width=0.95\linewidth]{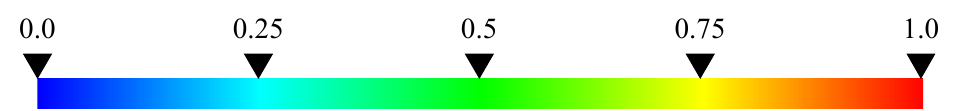}
  \caption{The similarity scale for the heatmap images}\label{fig:heatmap-scaler}
\end{figure}

%% file: inputs/experiment3/targets_with_compilers.tex
\newcommand{\tcmiddle}[1]{\multicolumn{1}{c}{\textbf{#1}}}

\begin{table}[tb]
    \centering
    \caption{The target software for the resilience evaluation}\label{table:target_exp3}
    {\footnotesize
    \begin{tabular}{l l l l r r}\hline \hline
\textbf{Arch.} & \textbf{OS} & \textbf{Compiler} & \textbf{Version} & \textbf{File size} & \textbf{\#Func.}\\ \hline
\multirow{6}{*}{\rotatebox{90}{amd64}} & \multirow{6}{*}{\rotatebox{90}{macOS}}
   & \bftt{clang}        & 21.0.0 &   115,768 &    63\\
&  & \bftt{gcc}          & 14.3.0 &   107,720 &    67\\
&  & \bftt{go}           & 1.26.3 & 3,221,776 & 2,559 \\
&  & \bftt{tinygo}       & 0.40.0 &   526,808 &   576 \\
&  & \bftt{rustc} (pure) & 1.95.0 & 2,420,576 & 3,413 \\
&  & \bftt{rustc} (lib)  & 1.95.0 & 2,409,496 & 3,369 \\ \hline
\multirow{6}{*}{\rotatebox{90}{amd64}} & \multirow{6}{*}{\rotatebox{90}{Linux}}
   & \bftt{clang}        & 14.0.0 &   252,936 &    68 \\
& & \bftt{gcc}          & 13.3.0 &   307,104 &    75 \\
& & \bftt{go}           & 1.26.3 & 3,293,291 & 2,572 \\
& & \bftt{tinygo}       & 0.40.0 & 1,706,784 &   701 \\
& & \bftt{rustc} (pure) & 1.95.0 & 2,180,616 & 3,387 \\
& & \bftt{rustc} (lib)  & 1.95.0 & 2,169,688 & 3,337 \\ \hline
\multirow{7}{*}{\rotatebox{90}{amd64}} & \multirow{7}{*}{\rotatebox{90}{Windows}}
  & \bftt{clang}        & 19.1.5 &   885,760 &  2,098 \\
& & \bftt{gcc}          & 15.2.0 &   359,898 &    111 \\
& & \bftt{msvc}         & 14.44  &   104,960 &    134 \\
& & \bftt{go}           & 1.26.3 & 3,277,312 &  2,326 \\
& & \bftt{tinygo} & \tcmiddle{---} & \tcmiddle{N/A} \\
& & \bftt{rustc} (pure) & 1.95.0 & 5,268,480 & 22,767 \\
& & \bftt{rustc} (lib)  & 1.95.0 & 5,237,248 & 22,232 \\ \hline
\multirow{6}{*}{\rotatebox{90}{arm64}} & \multirow{6}{*}{\rotatebox{90}{macOS}}
  & \bftt{clang}        & 21.0.0 &   129,240 &    64 \\
& & \bftt{gcc}          & 14.3.0 &   129,320 &    69 \\
& & \bftt{go}           & 1.26.3 & 3,095,906 & 2,518 \\
& & \bftt{tinygo}       & 0.40.0 &   546,352 &   989 \\
& & \bftt{rustc} (pure) & 1.95.0 & 2,267,184 & 3,336 \\
& & \bftt{rustc} (lib)  & 1.95.0 & 2,264,000 & 3,331 \\ \hline
\multirow{6}{*}{\rotatebox{90}{arm64}} & \multirow{6}{*}{\rotatebox{90}{Linux}}
  & \bftt{clang}        & 21.1.8 &   276,400 &    70 \\
& & \bftt{gcc}          & 15.2.0 &   335,912 &    83 \\
& & \bftt{go}           & 1.26.3 & 3,235,563 & 2,514 \\
& & \bftt{tinygo}       & 0.40.0 & 1,928,896 & 1,213 \\
& & \bftt{rustc} (pure) & 1.95.0 & 1,879,272 & 3,382 \\
& & \bftt{rustc} (lib)  & 1.95.0 & 1,859,128 & 3,331 \\ \hline
\multirow{7}{*}{\rotatebox{90}{arm64}} & \multirow{7}{*}{\rotatebox{90}{Windows}}
  & \bftt{clang}        & 19.1.5 &   664,576 &  2,597 \\
& & \bftt{gcc}          & 15.0.1 &   566,310 &    107 \\
& & \bftt{msvc}         & 14.44  &   101,376 &    130 \\
& & \bftt{go}           & 1.26.3 & 3,062,272 &  1,739 \\
& & \bftt{tinygo}  & \tcmiddle{---} & \tcmiddle{N/A} \\
& & \bftt{rustc} (pure) & 1.95.0 & 4,332,032 & 19,921 \\
& & \bftt{rustc} (lib)  & 1.95.0 & 4,304,896 & 19,480 \\ \hline
    \end{tabular}}
\end{table}

%% file: inputs/experiment3/scatter.tex
\begin{figure*}[bt]
  \begin{subcaptionblock}{0.31\linewidth}
    \includegraphics[width=\linewidth]{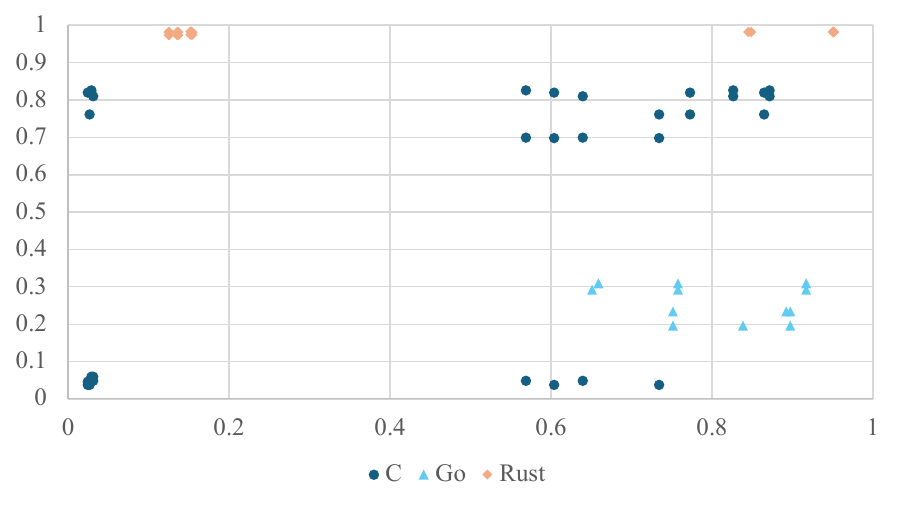}
    \caption{Dice}
  \end{subcaptionblock}
  \begin{subcaptionblock}{0.31\linewidth}
    \includegraphics[width=\linewidth]{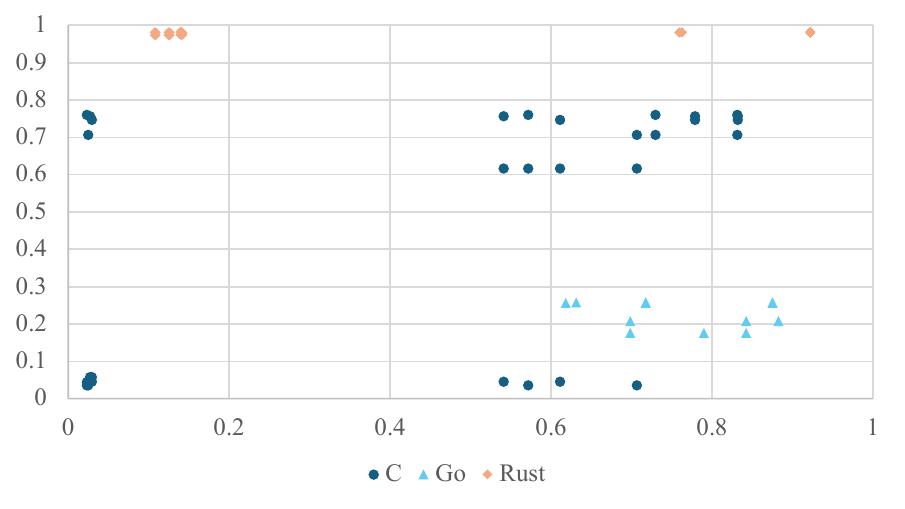}
    \caption{Jaccard}
  \end{subcaptionblock}
  \begin{subcaptionblock}{0.31\linewidth}
    \includegraphics[width=\linewidth]{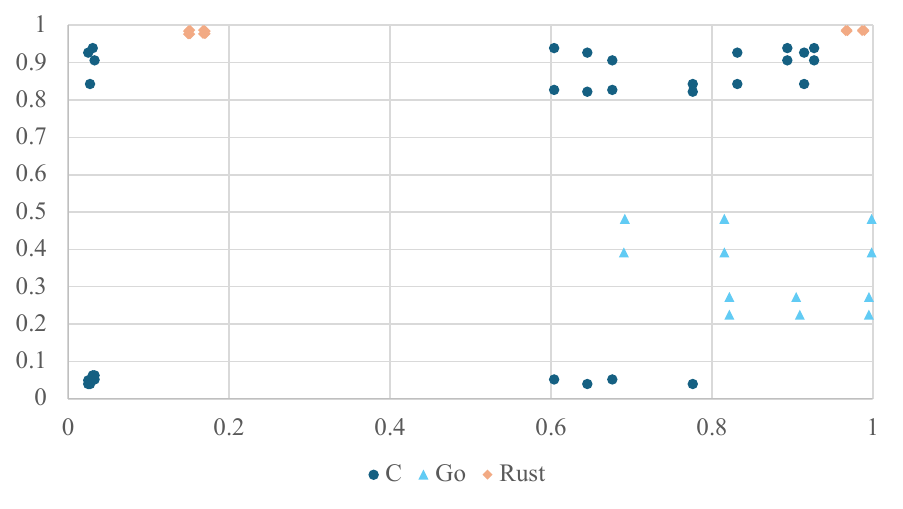}
    \caption{Simpson}
  \end{subcaptionblock}

  \vspace{0.1cm}

  \begin{subcaptionblock}{0.31\linewidth}
    \includegraphics[width=\linewidth]{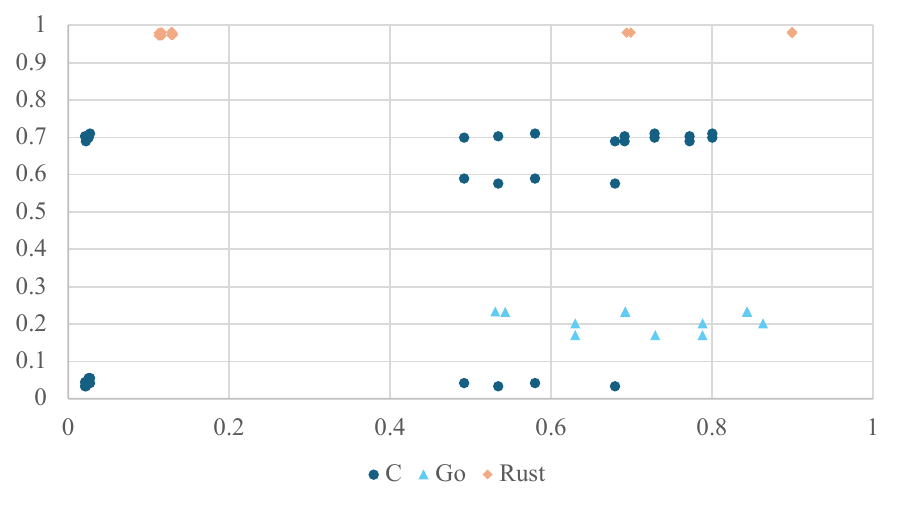}
    \caption{Euclidean}
  \end{subcaptionblock}
  \begin{subcaptionblock}{0.31\linewidth}
    \includegraphics[width=\linewidth]{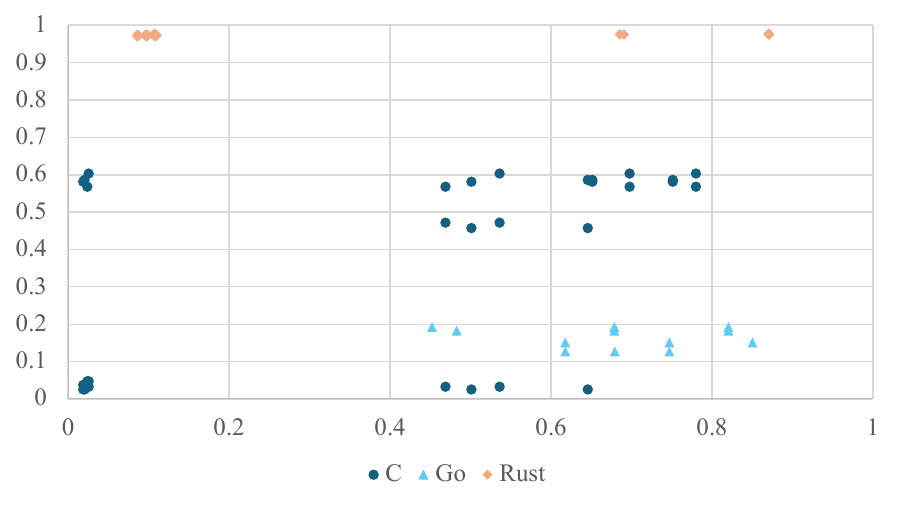}
    \caption{LCS}
  \end{subcaptionblock}
  \begin{subcaptionblock}{0.31\linewidth}
    \includegraphics[width=\linewidth]{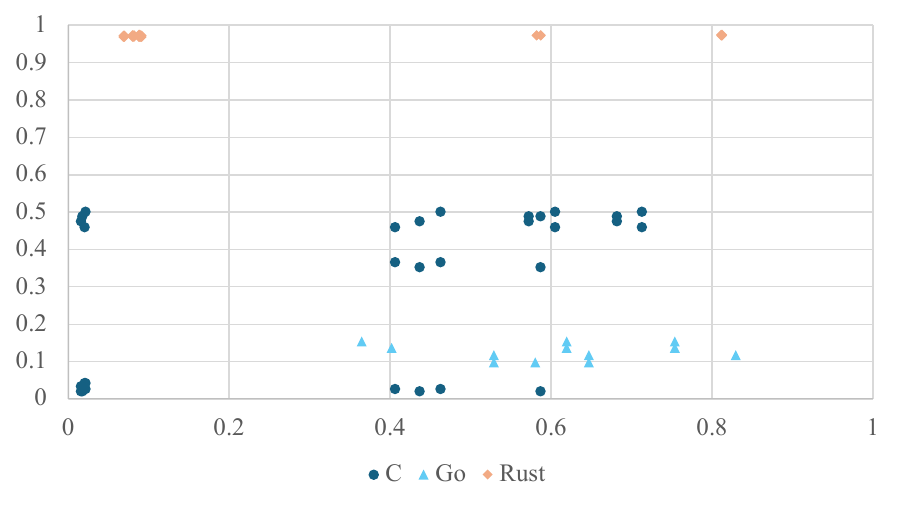}
    \caption{Levenshtein}
  \end{subcaptionblock}

  \vspace{0.1cm}

  \begin{subcaptionblock}{0.31\linewidth}
    \includegraphics[width=\linewidth]{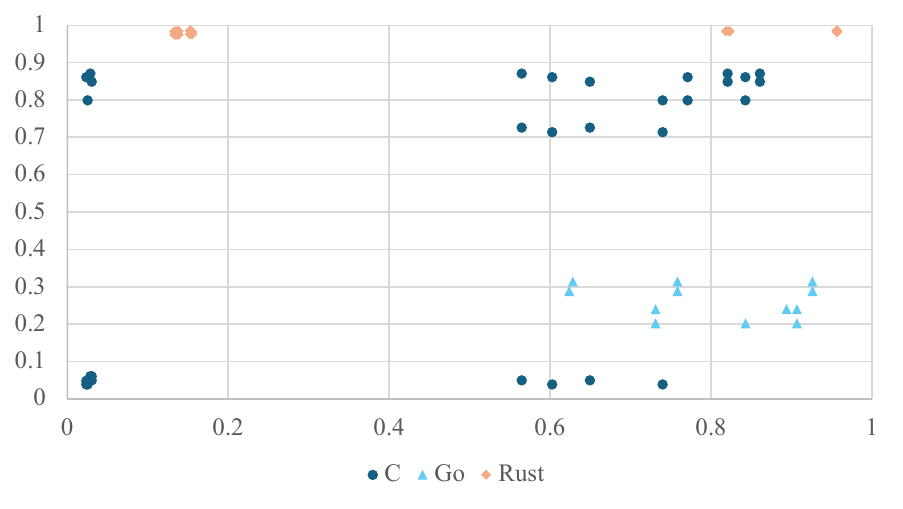}
    \caption{Cosine}
  \end{subcaptionblock}
  \begin{subcaptionblock}{0.31\linewidth}
    \includegraphics[width=\linewidth]{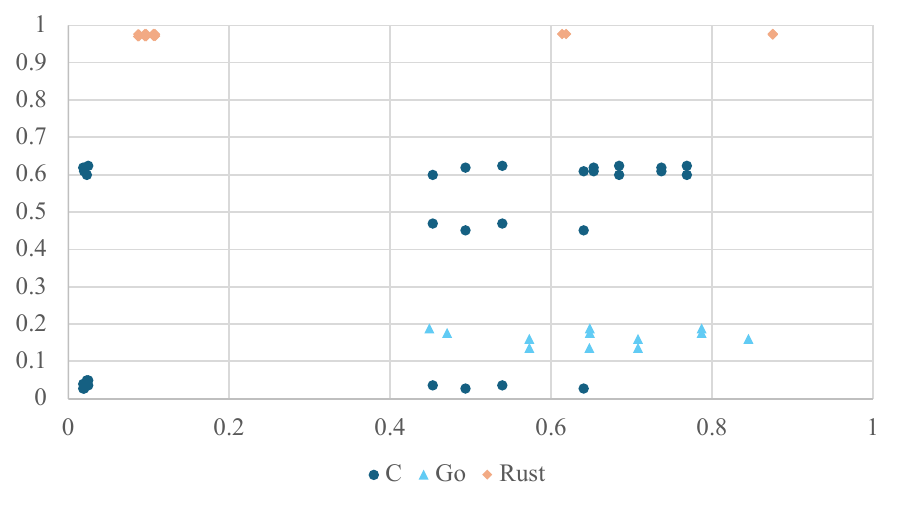}
    \caption{Weighted Jaccard}
  \end{subcaptionblock}

  \caption{Scatter plots of similarities across different operating systems and compilers.}\label{fig:resilience-scatter}
\end{figure*}